\documentclass[twocolumn,superscriptaddress,showpacs,preprintnumbers,amsmath,amssymb]{revtex4-1}
\usepackage{graphicx}
\usepackage{dcolumn}
\usepackage{bm}
\usepackage{float,color}
\usepackage{CJK}
\usepackage[colorlinks,linkcolor=blue,anchorcolor=blue,citecolor=blue]{hyperref}

\newcommand{\blue}{\textcolor[rgb]{0.00,0.00,1.00}}

\begin{document}
\preprint{preprint}
\title{A microscopic benchmark-study of triaxiality in low-lying states of $^{76}$Kr}%
\author{J. M. Yao}
\affiliation{School of Physical Science and Technology, Southwest University, Chongqing 400715, China}
\affiliation{Department of Physics, Tohoku University, Sendai 980-8578, Japan}
\author{K. Hagino}
\affiliation{Department of Physics, Tohoku University, Sendai 980-8578, Japan}
\affiliation{Research Center for Electron Photon Science, Tohoku University, 1-2-1 Mikamine, Sendai 982-0826, Japan}
\author{Z. P. Li}
\affiliation{School of Physical Science and Technology, Southwest University, Chongqing 400715, China}
\author{J. Meng}
\affiliation{State Key Laboratory of Nuclear Physics and Technology, School of Physics,
Peking University, Beijing 100871, China}
\affiliation{School of Physics and Nuclear Energy Engineering, Beihang University, Beijing 100191, China}
\affiliation{Department of Physics, University of Stellenbosch, Stellenbosch, South Africa}
\author{P. Ring}
\affiliation{State Key Laboratory of Nuclear Physics and Technology, School of Physics,
Peking University, Beijing 100871, China}
 \address{Physik-Department der Technischen Universit\"at M\"unchen, D-85748
         Garching, Germany}

\begin{abstract}
We report on a seven-dimensional generator coordinate calculation in the two deformation parameters $\beta$ and $\gamma$ together with projection on three-dimensional angular momentum and two particle numbers for the low-lying states in $^{76}$Kr. These calculations are based on covariant density functional theory. Excellent agreement is found with the data for the spectrum and the electric multipole transition strengths. This answers the important question of dynamic correlations and triaxiality in a fully microscopic way. We find that triaxial configurations dominate both the ground state and the quasi $\gamma$-band. This yields a different picture from the simple interpretation in terms of ``coexistence of a prolate ground state with an oblate low-lying excited state", which is based on the measured sign of spectroscopic quadrupole moments. This study also provides for the first time a benchmark for the collective Hamiltonian in five dimensions. Moreover, we point out that the staggering phase of the $\gamma$-band is not a safe signature for rigid triaxiality of the low-energy structure. 

\end{abstract}

\maketitle


\section{\label{sec:introduction}Introduction}

As has been disclosed by various spectroscopic methods that atomic nuclei, like other quantum many-body systems such as molecules, can display a variety of geometrical shapes. Changes of these shape are connected with collective motion. Most nuclei with proton and/or neutron open shells are axially deformed and characterized by the  quadrupole deformation parameter $\beta$. Some of them can have even non-axial shapes~\cite{Ring1980,Cwiok2005_Nature433-705}. In the past decades, there has been a growing interest in searching for deformed nuclei with triaxiality, which are described by the deformation parameters $\beta$ and $\gamma$. The existence of triaxial shapes in nuclei is of particular interest because it has provided a novel interpretation of many exotic phenomena, such as the violation of the $K$-selection rule in the decay of high-spin isomers~\cite{Chowdhury1988_NPA485-136}, the nuclear wobbling motion~\cite{Marshalek1979_NPA331-429,Odegaard2001_PRL86-5866}, and chiral rotations~\cite{Frauendorf1997_NPA617-131,Grodner2006_PRL97-172501,Meng2006PRC,Ayangeakaa2013_PRL110-172504}. However, whether a nucleus has rigid triaxiality or $\gamma$-softness at low energies is full of controversy. Since triaxiality cannot be measured directly, most of the discussions are model-dependent~\cite{Stachel1982_NPA383-429,Zamfir1991_PLB260-265,Sheikh1991_PLB507-115,Meng2010JPG}.

Recent measurements show evidence of triaxiality for some nuclei in $A\sim80$ mass region.  A typical example is the nucleus $^{76}$Kr. In calculations using collective Hamiltonians in five dimensions (5DCH) derived from energy density functionals (EDF), triaxiality turns out to be crucial to reproduce the spectroscopic properties of the low-lying states~\cite{Clement2007_PRC75-054313,Girod2009_PLB676-39,Sato2011_NPA849-53, FU-Y2013_PRC87-054305}. Later on, the odd-odd nucleus~$^{80}$Br shows a pattern of chiral vibration in the newly-measured excited states and is suggested to be a triaxial nucleus~\cite{WANG-SY2011_PLB703-40}. Most recently, $^{76}$Ge was pointed out to be a typical nucleus with a rigid triaxial deformation, because for its low-lying states the staggering behavior in the $\gamma$-band~\cite{Toh2013_PRC87-041304} is consistent with predictions of the rigid-triaxial rotor model of Davydov and Filippov (DF) with $\gamma=15^\circ$~\cite{Davydov1958_NP8-237}.

In the past decade, several EDF mapped approaches have been developed to study nuclear low-lying states with triaxiality. According to the most recent investigations~\cite{Nomura2012_PRC108-132501} based on the EDF mapped interacting boson model (IBM), neither the rigid-triaxial rotor model of DF~\cite{Davydov1958_NP8-237} nor the $\gamma$-unstable rotor model of Wilets and Jean (WJ)~\cite{Wilets1956_PRC102-788} is realized in actual nuclei. However, the EDF mapped IBM model is a semi-microscopic algebraic model in the sense that only the Hamiltonian is mapped to the energy surface derived from density functional theory (DFT). Microscopic mass parameters are not used for this mapping. The information of the underlying shell structure is not included. The 5DCH method to derive the potential of the collective Hamiltonian from the energy surface and the
mass parameters from the single-particle wave functions has turned out to be very successful for a microscopic investigation of spectroscopic data in nuclei both in non-relativistic~\cite{Libert1999_PRC60-054301,Prochniak2004_NPA730-59} and in relativistic~\cite{Prochniak2004_IJMPE13-217,Niksic2009_PRC79-034303,LI-Zhipan2009_PRC79-054301} density functional theories and for a global analysis of low-energy nuclear  spectroscopy~\cite{Bertsch2007_PRL99-032502,Delaroche2010_PRC81-014303}.

The form of the collective Bohr Hamiltonian has been derived in the literature from a microscopic Hamiltonian or from a microscopic density functional in two rather different ways, (i) from the generator coordinate method (GCM)~\cite{Haff1972_PRC7-951,Banerjee1973_ZPA258-46,Giraud1974_NPA233-373,Une1976_PTP55-498} and (ii) from time-dependent Hartree-Fock (TDHF) theory~\cite{Baranger1968_NPA122-241,Baranger1978_APNY114-123,Goeke1978_APNY112-328,Reinhard1987_RPP50-1}. In both cases additional approximations had to be used.

The derivation from the GCM method stays completely in the quantum mechanical framework. It relies on the validity of the Gaussian overlap approximation (GOA) for the overlaps between configurations with different deformations~\cite{LI-Zhipan2009_PRC79-054301} and on the assumption that the collective velocities are small, i.e. that the expansion in the collective momenta can be stopped after the second order.

In the derivation from TDHF theory the time-dependent densities are decomposed into generalized coordinates (time even parts) and momenta (time odd parts)~\cite{Baranger1978_APNY114-123}. In this case the TDHF equations have the form of classical equations of motion with a classical Hamiltonian function. In the adiabatic approximation, i.e. in adiabatic time-dependent Hartree-Fock (ATDHF) theory, one expands this function in terms of the momenta up to quadratic order. With the additional assumption, which is equivalent to the choice of the generator coordinates in the GCM ansatz, that there is a collective subspace decoupling from the other intrinsic degrees of freedoms, one neglects coupling terms and is left with a classical Bohr Hamiltonian in the collective coordinates. In the final step this function has to be quantized and one obtains a Bohr-Hamiltonian in the collective degrees of freedom.

Up to small details for the zero-point corrections to the potential energy there is an essential difference between the two methods in the kinetic terms. The inertia parameters derived from the GCM method correspond to the Peiers-Yoccoz inertia~\cite{Peierls1957_ProcPhysSocA70-381,Ring1980} and the inertia parameters derived from the ATDHF method correspond to the Thouless-Valatin inertia~\cite{Thouless1962_NP31-211}. In the case of translational motion, because of Galilean invariance, the proper inertia is the total mass $M=Am$ of the nucleus. It turns out that the Thouless-Valatin mass fulfills this condition, but the Peierls-Yoccoz inertia does not. The origin of this failure can be traced back to the fact, that in the conventional GCM-method one integrates in the Hill-Wheeler integral~\cite{Hill1953_RP89-1102,Griffin1957_PR108-311} only over the collective coordinates $q_i$, i.e. the Slater determinants $|q\rangle$ are time-even functions. If one uses an extended GCM-method and integrates as well over the coordinates $q_i$ as over the corresponding momenta $p_i$ one finds the proper value $M=Am$ for the inertia~\cite{Peierls1962_NP38-154}. It is obvious, that with present computer power this extended GCM method cannot be applied. Therefore it is generally assumed that one should use in the Bohr Hamiltonian the Thouless-Valatin inertia parameters. On the other side the full evaluation of these parameters is very complicated too. It basically requires the solution of the linear response equation and an inversion of the RPA-matrix at each point on the energy surface~\cite{Matsuyanagi10,LI-Zhipan2012_PRC86-034334}.

Therefore in most of the realistic applications an additional approximation is used, the residual interaction is neglected in the linear response equation. In this case one ends up with the well known Inglis-Belyaev formula for the rotational inertia and with a similar expression for the inertia in the vibrational degrees of freedom~\cite{Ring1980}. These parameters are usually called the {\it cranking inertia} or {\it cranking mass parameters}. In most applications of 5DCH based on cranking inertia and mass parameters the calculated energy of first excited 2$^+_1$ state is usually too large, i.e. the rotational moment of inertia is too small. Therefore in most of these applications the calculated spectrum is rescaled by a factor $\alpha\approx 1.4$~\cite{Libert1999_PRC60-054301}. The origin of this discrepancy  can be traced back to the fact that the residual interaction
in the denominator of the Thouless-Valatin inertia is on average attractive and therefore the denominator is reduced. It has been shown in Ref.~\cite{LI-Zhipan2012_PRC86-034334} that the rotational moment of inertia is increased by a factor $1.3-1.4$, when the residual interaction is fully taken into account.

In the recent years, the adiabatic selfconsistent collective coordinate (adiabatic SCC)  method has been proposed to derive the 5DCH~\cite{Matsuo00,Hinohara07}, where the equations of the SCC method~\cite{Marumori80} are solved using an expansion with respect to the collective momentum. In this method, both the vibrational and rotational collective masses were determined by local normal modes built on constrained HFB states. It has been shown that the time-odd components of the moving mean-field significantly increase the vibrational and rotational collective masses in comparison with the Inglis-Belyaev cranking masses~\cite{Hinohara09,Hinohara10,Sato2011_NPA849-53,Hinohara12}. However, these studies are carried out using a schematic pairing-plus-quadrupole Hamiltonian within several major-shell active model spaces both for neutrons and protons. A calculation with a modern energy functional has still been awaited.

From these considerations it is evident and this has also been pointed out in Ref.~\cite{FU-Y2013_PRC87-054305}, that the conclusions drawn from 5DCH calculations on the triaxiality in $^{76}$Kr might be different from those based on full projected GCM calculations in the coordinates $\beta$ and $\gamma$, which we will call, for the sake of simplicity, in the following ``full GCM calculations". To address the role of triaxiality and to understand the origin of simple patterns in the low-energy spectra of complex nuclei, in this work, we report on the first full microscopic calculation with triaxiality for the low-lying states in $^{76}$Kr, which also provides the first benchmark for the previous algebraic or geometrical model calculations.

The paper is arranged as follows: In Sec. \ref{sec:CGM-method} we give a short overview of the theoretical methods used in this work. Numerical details are discussed in Sec.~\ref{sec:Numerics} and the various energy surfaces are presented in Sec. \ref{sec:E-surfaces}. Results of the GCM calculations are compared with those of the 5DCH method in Sec.~\ref{sec:comparison} and Sec.~\ref{sec:conclusion} contains conclusions of these investigations.

\section{\label{sec:CGM-method} \blue{The Theoretical methods}}
The method used in this investigations is an extension of the beyond relativistic mean-field (RMF) approach presented in Refs.~\cite{Yao2010_PRC81-044311,Yao2011_PRC83-014308}. The starting point is provided by fully self-consistent constrained RMF+BCS calculations. The constrained quantities are the mass quadrupole moments $\langle Q_{20}\rangle$ and $\langle Q_{22}\rangle$ related to the triaxial parameters $\beta$ and $\gamma$.  All the mean-field triaxial states are subsequently projected onto designed particle numbers ($N, Z$) and angular momentum ($J$) by introducing the techniques of both particle number projection (PNP) and three-dimensional angular momentum projection (3DAMP)~\cite{Yao2008_CPL25-3609,Yao2009_PRC79-044312}. In the GCM method the quadrupole fluctuations about the mean-field solution are determined variationally by mixing all the projected states in the Hill-Wheeler integral~\cite{Hill1953_RP89-1102}. This level of implementation is also referred to as multi-reference (MR) DFT, which has become a standard and the state of the art microscopic model for studying nuclear low-lying collective excitations~\cite{Ring1980,Niksic2011_PPNP66-519}.

The nuclear many-body wave function is given as a linear combination of projected mean-field configurations generated by the collective coordinates of quadrupole deformations
\begin{equation}
 \label{TrialWF}
 \vert JNZ;\alpha\rangle
 =\sum_{q, K} f^{JK}_\alpha(q) \hat P^J_{MK} \hat P^N\hat P^Z\vert q \rangle.
 \end{equation}
where $\alpha=1,2,\ldots$ distinguishes different collective states with the same angular momentum $J$, and $\vert q\rangle=\vert \beta,\gamma\rangle$ denotes a set of RMF+BCS states with deformation parameters $(\beta,\gamma)$.  The operators $\hat{P}^{N}$, $\hat{P}^{Z}$, and $\hat P^J_{MK}$ project onto good neutron and proton numbers and onto good angular momentum. The weight coefficients $f^{JK}_\alpha(q)$ are determined by solving the Hill-Wheeler-Griffin equations~\cite{Hill1953_RP89-1102,Griffin1957_PR108-311} that are deduced from the minimization of the energy calculated with the GCM wave function~(\ref{TrialWF}). The solution of these equations provides the energy levels and all the information needed for calculating the electric multipole transition strengths. More details about the calculation of observables within this framework can be found in Ref.~\cite{Yao2010_PRC81-044311}. We note that two similar methods of MR-DFT (GCM+PN3DAMP) for triaxial nuclei have been developed recently in the non-relativistic scheme~\cite{Bender2008_PRC78-024309,Rodriguez2010_PRC81-064323}.

To provide a benchmark for the 5DCH method, which has been widely adopted for nuclear low-lying states, we also carry out the 5DCH calculation based on the same relativistic EDF and make a detailed comparison with the full GCM calculation. The collective Hamiltonian that describes the nuclear excitations of quadrupole vibration, rotation, and their couplings can be written in the form~~\cite{Libert1999_PRC60-054301,Prochniak2004_NPA730-59,Prochniak2004_IJMPE13-217,Niksic2009_PRC79-034303,LI-Zhipan2009_PRC79-054301}
\begin{equation}
\label{hamiltonian-quant}
\hat{H} =
\hat{T}_{\textnormal{vib}}+\hat{T}_{\textnormal{rot}}
              +V_{\textnormal{coll}} \; ,
\end{equation}
where $V_{\textnormal{coll}}$ is the collective potential that is given by the nuclear total energy corrected with the zero-point motions of rotation and vibration~\cite{Niksic2009_PRC79-034303}. The vibrational kinetic energy reads,
\begin{eqnarray}
\hat{T}_{\textnormal{vib}}
 &=&-\frac{\hbar^2}{2\sqrt{wr}}
   \left\{\frac{1}{\beta^4}
   \left[\frac{\partial}{\partial\beta}\sqrt{\frac{r}{w}}\beta^4
   B_{\gamma\gamma} \frac{\partial}{\partial\beta}\right.\right.\nonumber\\
  && \left.\left.- \frac{\partial}{\partial\beta}\sqrt{\frac{r}{w}}\beta^3
   B_{\beta\gamma}\frac{\partial}{\partial\gamma}
   \right]+\frac{1}{\beta\sin{3\gamma}} \left[
   -\frac{\partial}{\partial\gamma} \right.\right.\nonumber\\
  && \left.\left.\sqrt{\frac{r}{w}}\sin{3\gamma}
      B_{\beta \gamma}\frac{\partial}{\partial\beta}
    +\frac{1}{\beta}\frac{\partial}{\partial\gamma} \sqrt{\frac{r}{w}}\sin{3\gamma}
      B_{\beta \beta}\frac{\partial}{\partial\gamma}
   \right]\right\},
 \end{eqnarray}
and rotational kinetic energy,
\begin{equation}
\hat{T}_{\textnormal{\textnormal{\textnormal{rot}}}} =
\frac{1}{2}\sum_{k=1}^3{\frac{\hat{J}^2_k}{\mathcal{I}_k}},
\end{equation}
with $\hat{J}_k$ denoting the components of the angular momentum in
the body-fixed frame of a nucleus. It is noted that the mass
parameters $B_{\beta\beta}$, $B_{\beta\gamma}$, $B_{\gamma\gamma}$,
as well as the moments of inertia $\mathcal{I}_k$, depend on the
quadrupole deformation variables $\beta$ and $\gamma$.
Two additional quantities that appear in the expression for the
vibrational energy: $r=B_1B_2B_3$, and
$w=B_{\beta\beta}B_{\gamma\gamma}-B_{\beta\gamma}^2 $, determine the
volume element in the collective space. The corresponding eigenvalue
problem is solved using an expansion of eigenfunctions in terms of a
complete set of basis functions that depend on the deformation
variables $\beta$ and $\gamma$, and the Euler angles $\phi$,
$\theta$ and $\psi$. The dynamics of the 5DCH is governed by the seven
functions of the intrinsic deformations $\beta$ and $\gamma$: the
collective potential $V_{\rm coll}$, the three mass parameters:
$B_{\beta\beta}$, $B_{\beta\gamma}$, $B_{\gamma\gamma}$, and the
three moments of inertia $\mathcal{I}_k$, which are determined
by the single-(quasi)particle energies and wave functions from the mean-field calculations with the {\em cranking approximation}~\cite{Niksic2009_PRC79-034303,LI-Zhipan2009_PRC79-054301}. We point out here that we do not introduce a scaling factor for the moments of inertia in the 5DCH calculations, in contrast to previous studies in which the scaling factor has often been introduced in order to reproduce the energy of the first $2^+$ state.

\section{\label{sec:Numerics}Numerical details}

In the constrained triaxial RMF calculations, parity, $x$-simplex symmetry, and time-reversal invariance are imposed. The relativistic Kohn-Sham equation, i.e. the Dirac equation, is solved by expanding the Dirac spinors, separately for large and small components, in the basis of eigenfunctions of a three-dimensional harmonic oscillator in Cartesian coordinates~\cite{Koepf1989_NPA493-61,PENG2008_PRC78-024313,Niksic2014_CPC}. 10 major shells are found to be sufficient for the nuclei under consideration. We use the density functional derived from the relativistic point-coupling Lagrangian PC-PK1~\cite{ZHAO-PW2010_PRC82-054319}, for which dynamic correlation energies were found to improve its description of nuclear binding energy~\cite{ZhangQS2014FoP}. Pairing correlations between the nucleons are treated within the BCS approximation using a density-independent $\delta$-force with a smooth pairing window~\cite{Krieger1990_NPA517-275}. The strength parameters of the pairing force are $V_n=-349.5$ and $V_p=-330$ MeV$\cdot$fm$^3$ for neutrons and protons, respectively. The details of the pairing window are the same discussed in Ref.~\cite{Yao2010_PRC81-044311}. The Gauss-Legendre quadrature is used for the integrals over the three Euler angles ($\phi, \theta, \psi$) and the gauge angles $\varphi_{\tau=n,p}$ in the calculations of the projected kernels. The numbers of mesh-point in the interval $[0, \pi]$ are chosen as $(N_\phi=10, N_\theta=14, N_\psi=12)$ and $N_\varphi=9$. This turns out to be sufficient for the states with angular momentum $J\leq 6\hbar$~\cite{Yao2011_PRC84-024306}.  For the GCM calculations we have taken $N_{\rm GCM}=36$ relevant intrinsic states in the $(\beta, \gamma)$ plane.  The convergence of this calculation is checked by increasing or decreasing some configurations and examining the behavior of the collective wave functions and the energy dispersions~\cite{Bonche1990_NPA510-466}. The Paffian method~\cite{Robledo2009_PRC79-021302,Ballestero2011CPC} is implemented to calculate norm overlaps, the phase of which can be uniquely determined in this way. We note that the full projected GCM calculations on top of the covariant DFT are very time-consuming. Compared with the other two similar methods~\cite{Bender2008_PRC78-024309,Rodriguez2010_PRC81-064323}, the four-component Dirac spinors instead of two-component Pauli spinors are used for the single-particle wave functions, which makes the computational effort more demanding than in the non-relativistic cases. Specifically, computing each matrix element in the collective coordinate space takes about 200 CPU hours with one processor of 2.5 GHz Intel Xeon E5-2640. Parallelization techniques are utilized to reduce the computing time.

\section{\label{sec:E-surfaces}Energy surfaces}

\begin{figure*}[]
\centering
\includegraphics[width=18cm]{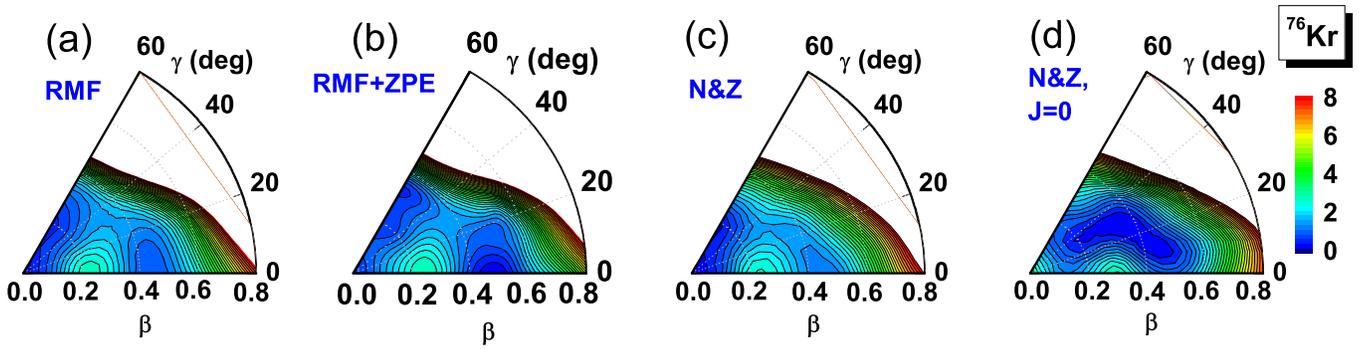}
\caption{\label{pes}(Color online)  Potential energy surfaces in the $(\beta,\gamma)$ plane for $^{76}$Kr:
(a) constrained RMF+BCS, (b) RMF+BCS with zero-point corrections, (c) with PNP, and (d) with both PNP and 3DAMP ($J=0$).
Two neighboring contour lines are separated by 0.25~MeV.}
\end{figure*}

Fig.~\ref{pes} displays various potential energy surfaces in the $(\beta, \gamma)$ plane for $^{76}$Kr. Panel (a) shows the mean field energy surface obtained by the constrained RMF+BCS method. A spherical minimum in the energy surface is found, soft along oblate shapes and competing with a large prolate deformed minimum. In panel (b) collective potential $V_{\rm coll}$ with zero-point corrections used in the 5DCH method is shown. Details are discussed in Ref.~\cite{Niksic2009_PRC79-034303}. These corrections do not change the energy surface in a qualitative way, but they lead to somewhat larger deformations of the minima and therefore to an enhanced collectivity. Panel (c) of Fig.~\ref{pes} displays the energy surfaces obtained from wave functions with exact particle number projection (PNP) after the variation and panel (d) with additional three-dimensional AMP. PNP alone does not lead to large deviations from the mean field surface. 3DAMP, however, changes the picture considerably. Of particular interest is here the onset of a triaxial minimum, soft along the direction connecting the weakly oblate deformed state (with $\vert\beta\vert\approx0.2$) and the strongly prolate deformed state (with $\vert\beta\vert\approx0.5$). Both of them become saddle points in the triaxial energy surface. As shown in Refs.~\cite{Bender2006_PRC74-024312,FU-Y2013_PRC87-054305}, the beyond mean-field calculation restricted to axial symmetry misinterpreted the weakly oblate deformed configuration as the ground state and therefore failed to reproduce the low-energy structure of the nucleus $^{76}$Kr. The failure of the previous studies without triaxiality can be understood from Fig.~\ref{pes}. Moreover, even though the previous studies with triaxiality for other nuclei have already shown that the dynamical correlation energy from restoration of rotational symmetry can lower the energy of triaxial states~\cite{Hayashi1984_PRL53-337,Bender2008_PRC78-024309,Yao2011_PRC84-024306}, the phenomenon presented here is obviously rare and very interesting in the sense that the energetically favored triaxial states connect oblate and prolate states of very different deformations, changing dramatically the topological structure of the energy surface. This phenomenon seems to be a particular feature of nuclei of this mass region. A similar phenomenon is also shown in $^{80}$Zr in the calculations of Ref.~\cite{Rodriguez2011_PLB705-255} but with the presence of several local triaxial minima. The description of low-energy states in nuclei with such complicated structures is definitely a challenge for a full microscopic model.

\section{\label{sec:comparison}Comparison of GCM and 5DCH results}

Since the EDF mapped 5DCH model has been extensively adopted to study nuclear low-lying states, it is interesting to make a detailed comparison between the full GCM calculation and the 5DCH calculation for $^{76}$Kr. The full GCM calculation can provide a benchmark for the EDF mapped 5DCH calculation.

\begin{figure*}[]
\centering
\includegraphics[width=19cm]{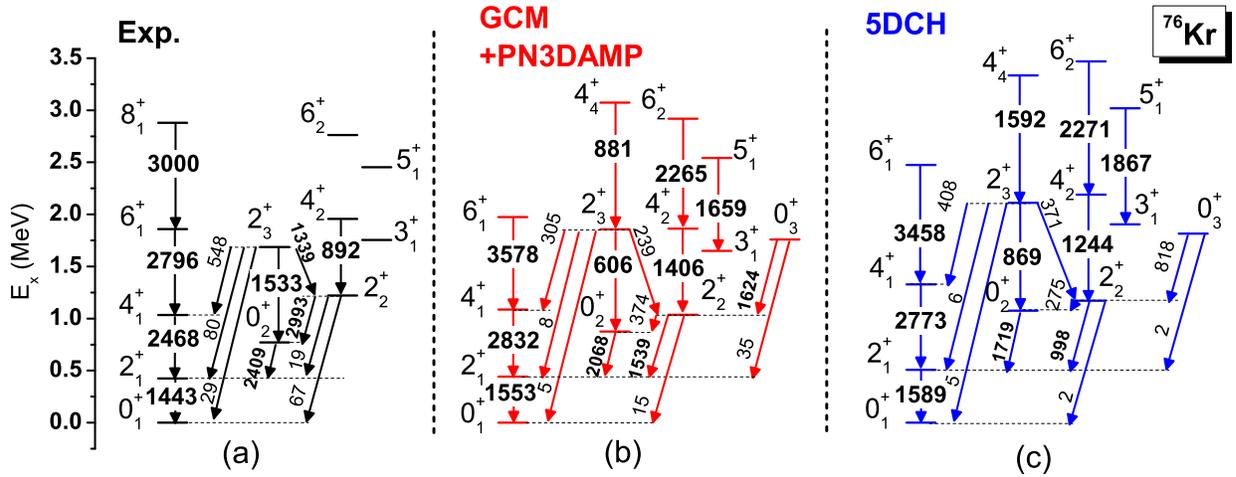}\vspace{-0.5cm}
\caption{\label{levels1}(Color online) Low-lying spectra and $B(E2)$ values (in $e^2$fm$^4$) of $^{76}$Kr.
Results from the full relativistic GCM calculation with PNP and 3DAMP (b) are compared with 5DCH results (cf. Fig.~9 of Ref.~\cite{FU-Y2013_PRC87-054305}) (c) and with experimental data (a) from Ref.~\cite{Clement2007_PRC75-054313}.}
\end{figure*}

\subsection{\label{sec:spectra}Spectra}
\begin{table}
\tabcolsep=3pt
\setlength{\extrarowheight}{4pt}
\caption{Energy levels  (in MeV) for low-lying states of $^{76}$Kr derived from triaxial relativistic GCM+PN3DAMP calculations and from 5DCH calculations are compared with data. Both calculations are base on the same EDF PC-PK1.}
\begin{tabular}{c|c|c|c}
  \hline
  \hline
    $J^\pi$      &  Exp. & GCM    & 5DCH \\
  \hline
  ~$0^+_1$~~&~~~$0.0$~~~~~~&~~~$0.0$~~~~~~&~~~$0.0$~~~~~\\
  $2^+_1$~~&~~~$0.424$~~~&~~~$0.441$~~~&~~~$0.508$  \\
  $4^+_1$~~&~~~$1.035$~~~&~~~$1.087$~~~&~~~$1.327$  \\
  $6^+_1$~~&~~~$1.859$~~~&~~~$1.975$~~~&~~~$2.474$  \\
  $8^+_1$~~&~~~$2.879$~~~&~~~       ~~~&~~~$3.936$  \\
\hline
  $0^+_2$~~&~~~$0.770$~~~&~~~$0.876$~~~&~~~$1.075$  \\
  $2^+_3$~~&~~~$1.688$~~~&~~~$1.857$~~~&~~~$2.111$  \\
  $4^+_4$~~&             &~~~$3.073$~~~&~~~$3.334$  \\
\hline
  $2^+_2$~~&~~~$1.222$~~~&~~~$1.036$~~~&~~~$1.171$  \\
  $3^+_1$~~&~~~$1.756$~~~&~~~$1.650$~~~&~~~$1.905$  \\
  $4^+_2$~~&~~~$1.957$~~~&~~~$1.864$~~~&~~~$2.190$  \\
  $5^+_1$~~&~~~$2.452$~~~&~~~$2.543$~~~&~~~$3.018$  \\
  $6^+_2$~~&~~~$2.763$~~~&~~~$2.920$~~~&~~~$3.470$  \\
  \hline
  $0^+_3$~~&~~~       ~~~&~~~$1.760$~~~&~~~$1.816$  \\
  \hline
\end{tabular}
\label{tab1}
\end{table}

Fig.~\ref{levels1} displays the low-lying spectra of $^{76}$Kr. All the calculations are based on the relativistic point-coupling Lagrangian PC-PK1~\cite{ZHAO-PW2010_PRC82-054319}. The full GCM calculation with number projection and three-dimensional angular momentum projection is compared with the experimental data and with results of the 5DCH calculation, cf. Fig.~9 in Ref.~\cite{FU-Y2013_PRC87-054305}, where we have presented two 5DCH results (Fig.6 and Fig.9 in Ref.~\cite{FU-Y2013_PRC87-054305})  for $^{76}$Kr using two different pairing forces. To make a comparison with the present GCM calculation, only the results by the same pairing force as adopted here are plotted in Fig.~\ref{levels1}(c). Both GCM and 5DCH calculations yield similar structures, a ground state rotational band, a quasi-$\beta$-band with a band head $I^\pi=0^+_2$ and a quasi-$\gamma$-band with a band head $I^\pi=2^+_2$. The results are in good agreement with the data. In detail, this agreement is excellent for the full GCM-calculation and clearly superior to that for the 5DCH result. In particular, the large electric quadrupole transition from the low-lying $0^+_2$ state of the quasi-$\beta$-band to the $2^+_1$ state in the ground state band is reproduced automatically and definitely better than in the 5DCH calculations. These results settle the important role of triaxiality, despite the fact that in the mean-field energy surface the triaxial states are only saddle points as shown in Fig.~\ref{pes}(a). PNP does not change the situation very much, we still have a ridge of roughly 0.5 MeV between the two axial minima. Only in the case of 3DAMP (Fig.~\ref{pes}(d))  this ridge disappears and a shallow triaxial minimum with a depth of roughly 0.5 MeV develops. The 5DCH method uses the unprojected energy surface with zero-point corrections (Fig.~\ref{pes}(b)). The fact that its spectrum is still rather close to the full GCM spectrum with 3DAMP must therefore depend on the behavior of the mass parameters. All these observations indicate that the role of triaxiality in nuclear low-lying states cannot be justified simply on the basis of the mean-field energy surface. As usual the 5DCH spectrum is stretched as compared to the data. As we see from the $2^+_1$ level in Table.~\ref{tab1}, the stretching factor is 1.2, slightly smaller than the usual factor 1.4. but it increases with increasing spin and for the $8^+_1$ we have already 1.38. On the other hand the GCM spectrum is very close to the experiment. From the $2^+_1$ level we derive a very small stretching factor 1.04 which stays roughly constant.

\begin{figure*}[]
\centering
\includegraphics[width=18cm]{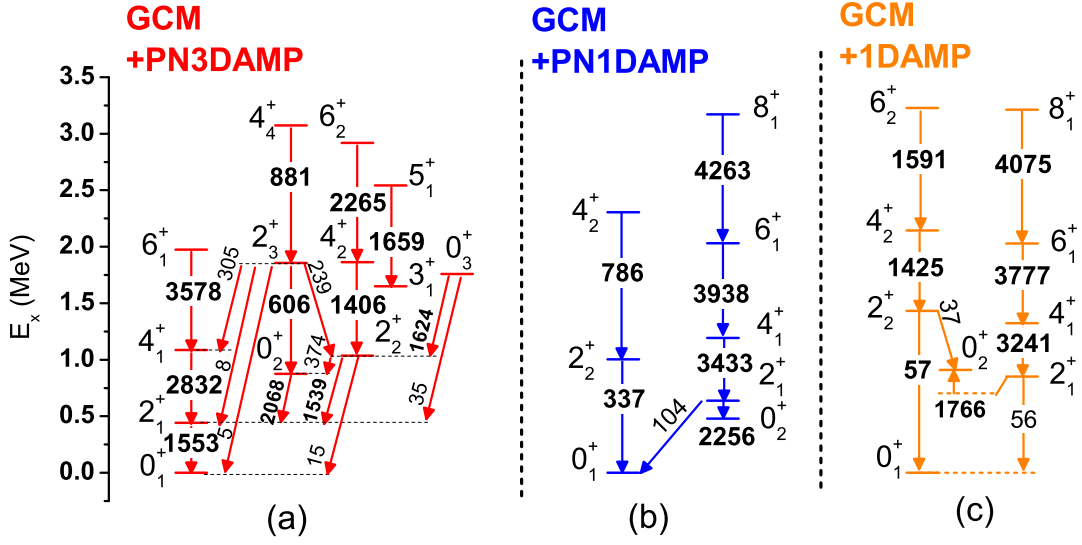}\vspace{-0.5cm}
\caption{\label{levels2}(Color online) Same as in Fig.~\ref{levels2}. The
results from triaxial relativistic GCM calculations with 3DAMP and PNP (a) are compared with axial calculations with 1DAMP and PNP (b) and with axial calculations with 1DAMP only (c).}
\end{figure*}

Fig.~\ref{levels2} shows a comparison between the full GCM+PN3DAMP calculation in the coordinates $\beta$ and $\gamma$ with axially symmetric calculations with the coordinate $\beta$ only, i.e. along the two lines  in Fig.~\ref{pes} with $\gamma=0$ (prolate deformations with $\beta>0$) and $\gamma=60^\circ$ (oblate deformations with $\beta<0$). It is evident that a restriction to axial states fails to reproduce the low-energy structures of the spectrum. As it has been found already in Refs~\cite{Bender2006_PRC74-024312,FU-Y2013_PRC87-054305} the axial calculation predicts the coexistence of a weakly oblate ground-state band with a prolate excited band. Fig.~\ref{levels2}(c) shows, for the axial case, a calculation without number projection. The lower part of the spectrum is very much disturbed in the case without number projection. It is evident that the band heads $0^+_1$ and $0^+_2$ are considerably shifted. This can be understood by the spurious mixing between the $0^+$-states in the nucleus under consideration with corresponding $0^+$-states in nuclei with different particle numbers $N{\pm}2$ or $Z{\pm}2$. This shows clearly that in cases of transitional nuclei with flat energy surfaces, changing considerably with the number of particles, conservation of particle number is definitely essential for a full understanding of the experimental data.
As discussed in Ref.~\cite{Yao2011_PRC83-014308}, it is not guaranteed that the wave functions of GCM calculations with only angular momentum projection have the correct average particle numbers. There are also unphysical interference terms. On the other side, it is hard to understand, why the 5DCH calculations work so well, because they are based on the unprojected potential energy surface. Here the fermionic degrees of freedom with individual particles are eliminated and particle number is not an issue. A better understanding requires further investigations in future.

\begin{figure}[]
\centering
\includegraphics[width=7cm]{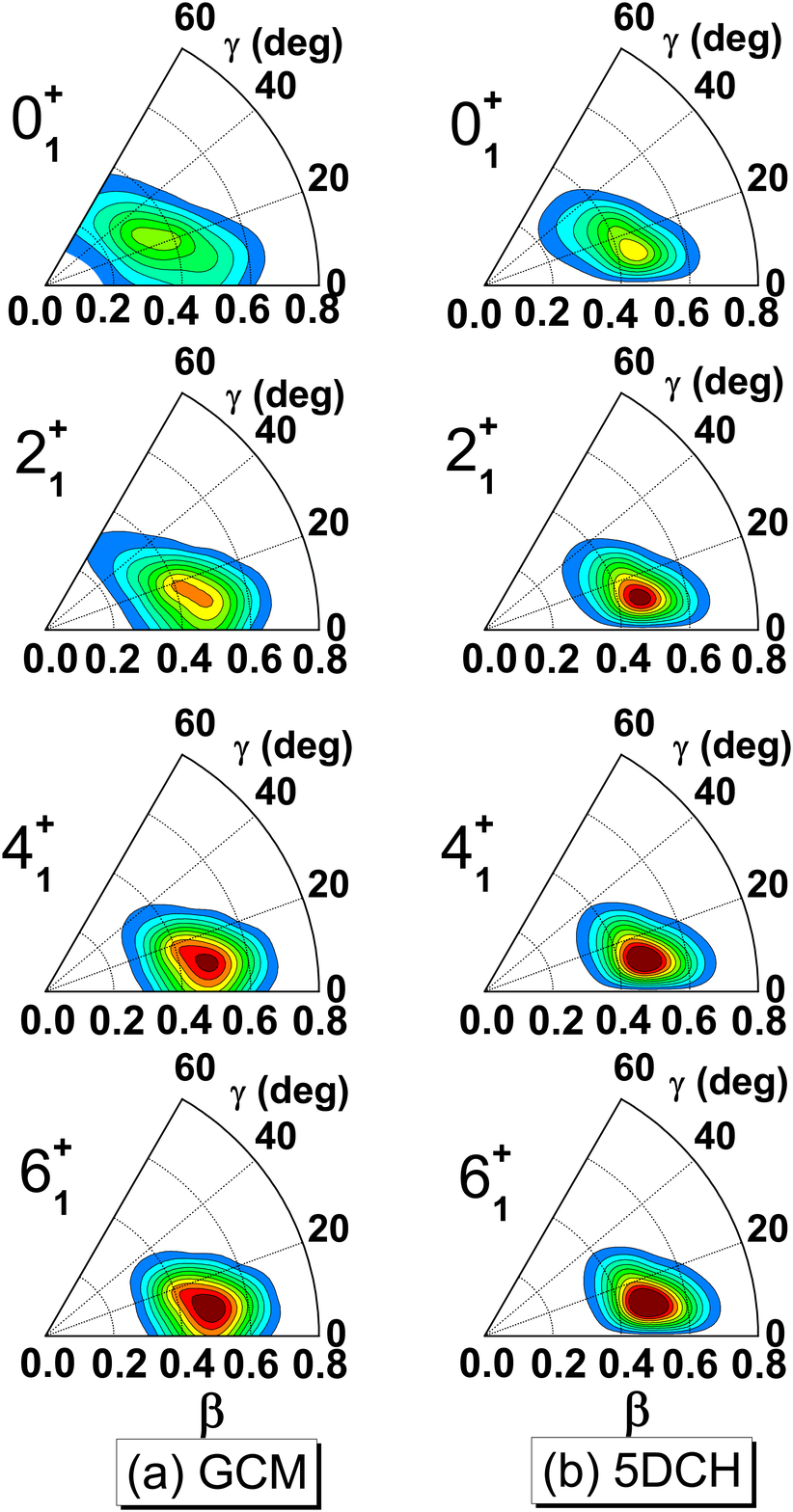}
\caption{\label{gsb}(Color online) Square of the collective wave functions in the $(\beta,\gamma)$ plane for the ground state band of $^{76}$Kr calculated (a) by full GCM+PN§DAMP and (b) by 5DCH.}
\end{figure}

\begin{figure}[]
\centering
\includegraphics[width=7cm]{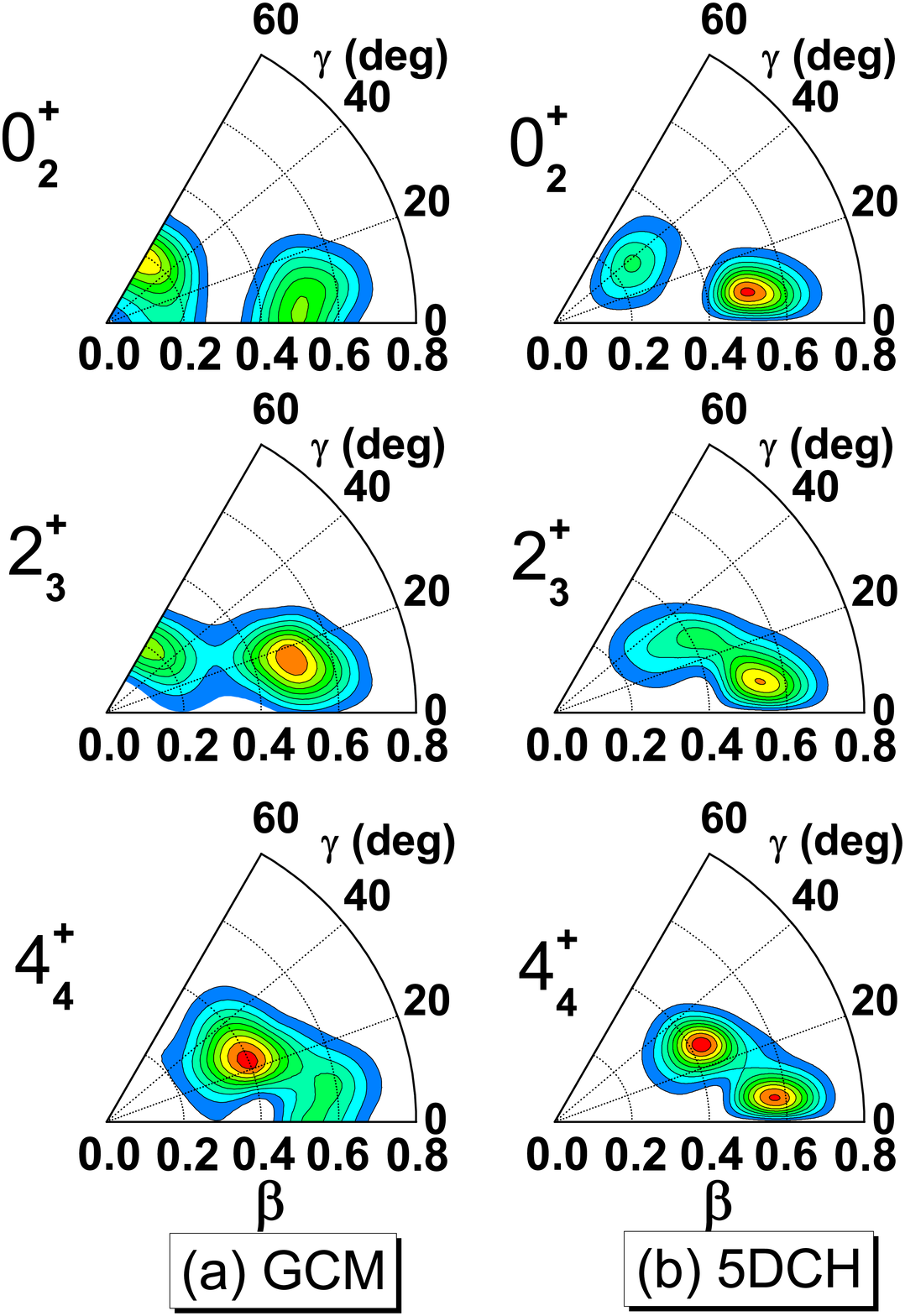}
\caption{\label{betab}(Color online) Same as in Fig.~\ref{gsb} but for the quasi-$\beta$-band calculated (a) by full GCM+PN§DAMP and (b) by 5DCH.}
\end{figure}

\begin{figure}[]
\centering
\includegraphics[width=7cm]{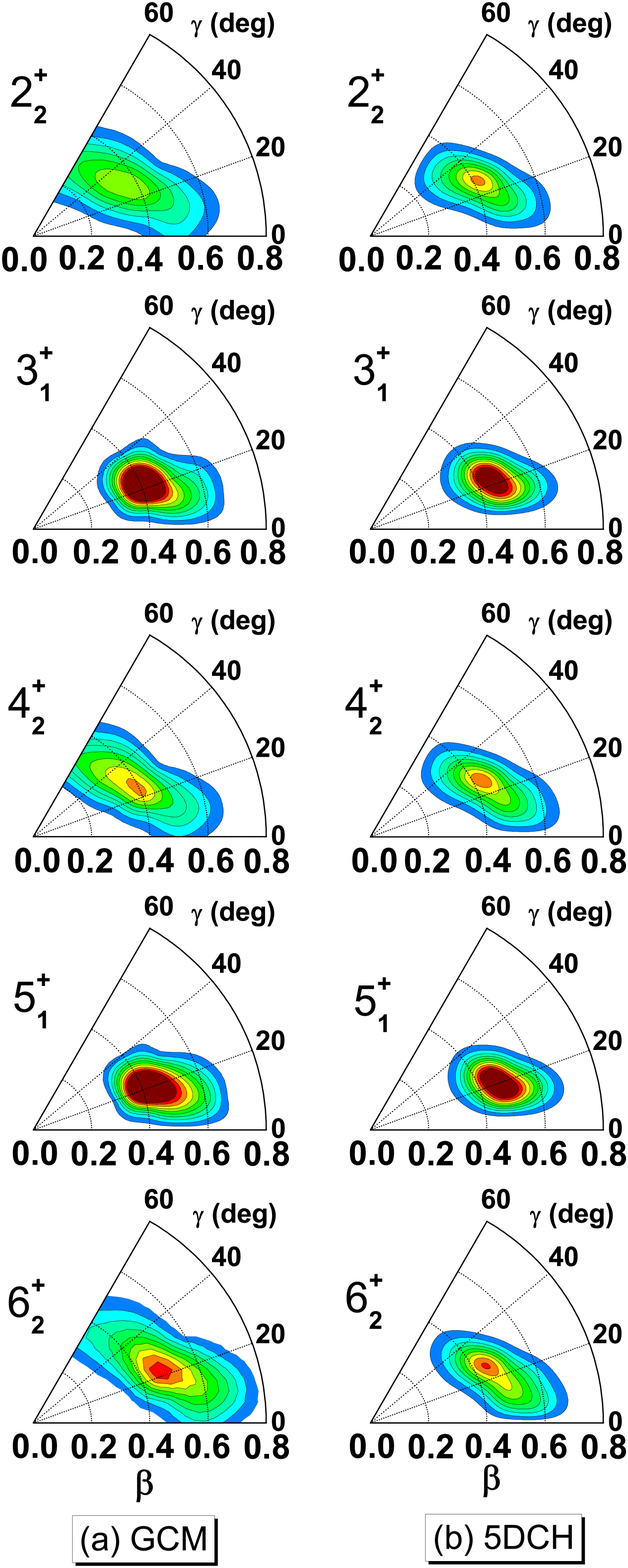}
\caption{\label{gamb}(Color online) Same as in Fig.~\ref{gsb} but for the quasi-$\gamma$-band calculated (a) by full GCM+PN§DAMP and (b) by 5DCH.}
\end{figure}

\subsection{\label{sec:wavefunctions}Collective wave functions}


Next we investigate the wave functions resulting from the two models. In Figs.~\ref{gsb}, \ref{betab}, and \ref{gamb} we compare the probability densities $\rho^{\rm GCM}_{J\alpha}(q)$ obtained from full GCM+PN3DAMP calculations,
\begin{equation}
\label{GCM_wf}
\rho^{\rm GCM}_{J\alpha}(q)
= \sum_K\Big\vert \sum_{q'K'} [ \mathcal{N}^{J}_{KK'}(q,q') ]^{1/2} f^{JK'}_\alpha(q')\Big\vert^2,
\end{equation}
 where the norm kernel is defined by $\mathcal{N}^{J}_{KK'}(q,q') =\langle q \vert \hat P^J_{KK'} \hat P^N\hat P^Z\vert q'\rangle$,  with the corresponding  probability densities $\rho^{\rm CH}_{J\alpha}(q)$ of the 5DCH results
 \begin{eqnarray}
\label{CH_wf}
   \rho^{\rm CH}_{J\alpha}(q) = \sum_K \Big\vert \psi^{JK}_\alpha  (q)\Big\vert^2  \beta^4 \sin3\gamma,
 \end{eqnarray}
 for the different bands shown in Fig.~\ref{levels1}. In Eq.~(\ref{CH_wf}), the $ \psi^{JK}_\alpha  (q)$ is the deformation-dependent part of the collective wave function in the 5DCH model~\cite{LI-Zhipan2009_PRC79-054301}. We note that the summation of the probability density in both Eq.(\ref{GCM_wf}) and Eq.(\ref{CH_wf}) over the entire collective coordinates $q(\beta, \gamma)$ is unity.

Fig.~\ref{gsb} shows the ground state band. Triaxial configurations dominate for all angular momenta. At the band head, i.e. at the ground state we find a relatively broad distribution extended rather far in $\gamma$ direction and concentrated on a certain range in $\beta$-values, in general agreement with the structure of the energy surfaces given in Fig.~\ref{pes}. This is in particular true for the full GCM calculations, where the PN3DAMP energy surface in Fig.~\ref{pes}(d) has a clear minimum at roughly $\beta\approx 0.4$ and $\gamma\approx 20^\circ$ and the wave function extends roughly over the full blue area in this figure. With increasing angular momentum the distribution in $\gamma$-direction becomes increasingly narrow. This has probably to do with the fact that the angular momentum is created by aligning partially certain single particle angular momenta in one direction and this favours certain deformations forming a spatial overlap with those configurations. In the 5DCH calculations the situation is very similar. The only difference occurs at angular momentum zero, where we find a wave function in 5DCH relatively concentrated around $\beta\approx 0.45$ and $\gamma\approx 14^\circ$. This can be understood qualitatively from the energy surface in Fig.~\ref{pes}(b), which forms the basis of the 5DCH calculations. Here we have a minimum in this $\beta$,$\gamma$ region and the ridge at $\gamma\approx 30^\circ$ forbids a much further extension in $\gamma$ direction. With increasing angular momentum we find a very close agrement for the wave functions in both models.
In both cases we find wave functions which are for increasing angular momentum more and more concentrated in a relatively narrow area in the $\beta$-$\gamma$ plane. This is interesting because the arguments given above for the microscopic GCM calculations do not apply directly to the Bohr model, which has for all angular momenta the same energy surface given in Fig.~\ref{pes}(b). The origin of this concentration of the wave function to one narrow area in the $\beta$-$\gamma$ plane must be caused by the kinetic part of the 5DCH, i.e. by the mass parameters. Of course this requires further investigations. A similar concentration of the wave functions in the present 5DCH calculation has also been reported in 5DCH calculation based on the quadrupole plus pairing model with the vibrational and rotational mass parameters determined by local QRPA calculations in Ref.~\cite{Sato2011_NPA849-53} for Kr isotopes and in Refs.~\cite{Hinohara09,Hinohara10} for Se isotopes. 

In Fig. \ref{betab} we show the probability distributions for the quasi-$\beta$-band. Again we have a close similarity between the full GCM-calculation and the 5DCH. At the band-head, the low-lying $0^+_2$ state contains a mixing of weakly oblate configurations and large prolate configurations, instead of a pure oblate state. This provides a different picture as compared to the interpretation of the ``coexistence of a prolate ground state with an oblate low-lying excited state" based on the measured sign of spectroscopic quadrupole moments in Ref.~\cite{Clement2007_PRC75-054313}. In particular for the GCM-calculations, the probability is concentrated mostly along the symmetry axis. It shows two peaks. It turns out that the sign of the corresponding wave function is different on both sides, i.e. this wave function has a node. In a simplified picture of a one-dimensional oscillator potential this would correspond to the first excited state of a vibration in $\beta$-direction. For the higher members of this band this is, however, no longer true. We observe a concentration of the probability in the triaxial regime.

The quasi-$\gamma$-band in Fig.~\ref{gamb} shows again a very similar behavior for the GCM and for the 5DCH calculations. For even angular momenta the probability is distributed over a narrow region of $\beta$-values and a rather wide region of $\gamma$-values in both cases. On the contrary the distributions for odd $I$-values are sharply peaked at $\beta\approx 0.4$ and $\gamma\approx 20^\circ$. This strong staggering is also observed in the spectrum in Fig.~\ref{levels2}.

\subsection{\label{sec:quadrupole}Quadrupole momenta}
\begin{table}
\tabcolsep=3pt
\setlength{\extrarowheight}{2pt}
\caption{Spectroscopic quadrupole moments $Q^{\rm s}$ (in $e$b) for low-lying states of $^{76}$Kr from triaxial relativistic GCM+PN3DAMP calculations and from 5DCH calculations based on both non-relativistic and relativistic EDFs, in comparison with data.}
\begin{tabular}{ccc|ccc}
  \hline
  \hline
   &   Exp. & GCM   & \multicolumn{3}{c}{5DCH} \\
  \hline

  $J^\pi$ & \cite{Clement2007_PRC75-054313} &   PC-PK1 &  PC-PK1~\cite{FU-Y2013_PRC87-054305} &  SLy6~\cite{FU-Y2013_PRC87-054305} & D1S~\cite{Clement2007_PRC75-054313} \\
  \hline
  $2^+_1$ & $-0.7\pm0.2$ & $-0.61$ & $-0.71$ &$-0.72$ & $-0.50$ \\
  $4^+_1$ & $-1.7\pm0.3$ & $-0.95$ & $-0.99$ &$-1.02$ & $-0.85$ \\
  $6^+_1$ & $-2.0\pm0.3$ & $-1.07$ & $-1.11$ &$-1.16$ & $-1.01$ \\
  $2^+_3$ & $+1.0\pm0.4$ & $+0.46$ & $-0.19$ &$-0.29$ & $+0.04$ \\
  $2^+_2$ & $-0.7\pm0.3$ & $+0.36$ & $+0.50$ &$+0.48$ & $+0.26$ \\
  $3^+_1$ &              & $+0.00$ & $+0.00$   & $+0.00$ &        \\
  $4^+_2$ &              & $+0.02$ & $-0.17$ &$-0.19$ &        \\
  $5^+_1$ &              & $-0.57$ & $-0.57$ &$-0.58$ &       \\
  $6^+_2$ &              & $-0.45$ & $-0.60$ &$-0.59$ &      \\
  \hline
  \hline
\end{tabular}
\label{tab2}
\end{table}

In Table~\ref{tab2}, we give the spectroscopic quadrupole moments ($Q^{\rm s}$) from both calculations.  The $Q^{\rm s}$ values of the full GCM calculations are close to those of the 5DCH calculations based on the same relativistic EDF for most low-lying states except for the high-lying $2^+_3$ and $4^+_2$ states. We note that the 5DCH results, including the spectra and $B(E2)$ values are not very sensitive to the underlying EDF's. Both the non-relativistic EDF's and the relativistic EDF give similar results. Moreover, all the calculations predict an opposite sign of $Q^{\rm s}$ to the data for the $2^+_2$ state.

\begin{figure}[]
\centering
\includegraphics[width=8.5cm]{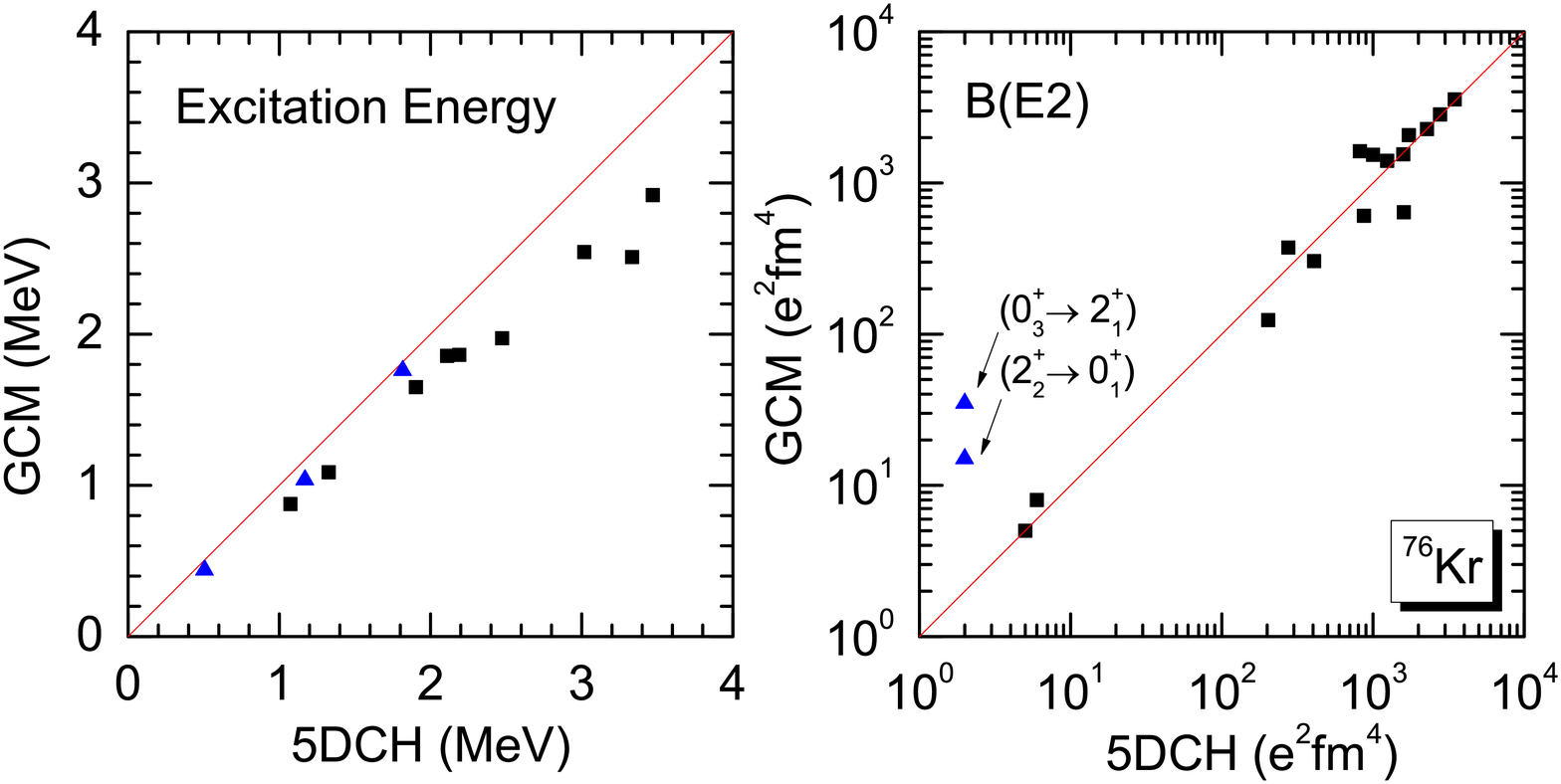}
\caption{\label{comparison}(Color online) Comparison between the results of full GCM and the 5DCH calculations for the low-lying states in $^{76}$Kr. The same underlying EDF (PC-PK1) is used. }
\end{figure}

In Fig.~\ref{comparison} a detailed comparison is made between full GCM and 5DCH results for the excitation energies and the electric quadrupole transition strengths. It is seen that, as discussed before, the 5DCH method produces excitation energies systematically higher than those of the full GCM calculation by $\sim20\%$. Nevertheless, the $B(E2)$ values obtained by the 5DCH method agree better with the full GCM results, except for the weak out-of-band $E2$ transitions $0^+_3\to2^+_1$ and $2^+_2\to0^+_1$.  We note that the mass parameters here are calculated using the perturbative cranking expression, which leads to systematically larger values than those of Gaussian overlap approximation (GOA), as demonstrated in Ref.~\cite{Baran2011_PRC84-054321}. In other words, the excitation energies would be overestimated further if the GOA mass parameters are used in the 5DCH calculations.  One promising way to achieve an improved description of excitation energies in 5DCH calculations is staying within the ATDHF approximation and using the Thouless-Valatin inertia, which can be obtained either by a selfconsistent cranking calculation using a very small cranking frequency~\cite{Delaroche2010_PRC81-014303} or by the method based the rapid convergence of the expansion of the inertia matrix~\cite{LI-Zhipan2012_PRC86-034334}. Of course, the ATDHF approach does not justify the zero-point energy corrections for the potential and the lowering of the energy by additional correlations. On the other side, as discussed in Ref.~\cite{Ring1980}, these two parts cancel each other to a large extend and are therefore not very essential. The Thouless-Valatin inertia can, in principle, also be derived from an extended GCM-method including not only the collective coordinates $q$ (time-even components) but also the corresponding momenta $p$ (time-odd parts). This is equivalent to the use of complex generator coordinates. Of course, present computational facilities do not allow applications of such extended methods in the framework of realistic density functional theories.


\subsection{\label{sec:staggering}Staggering behavior of the $\gamma$-band.}

\begin{figure}[]
\centering
\includegraphics[width=7cm]{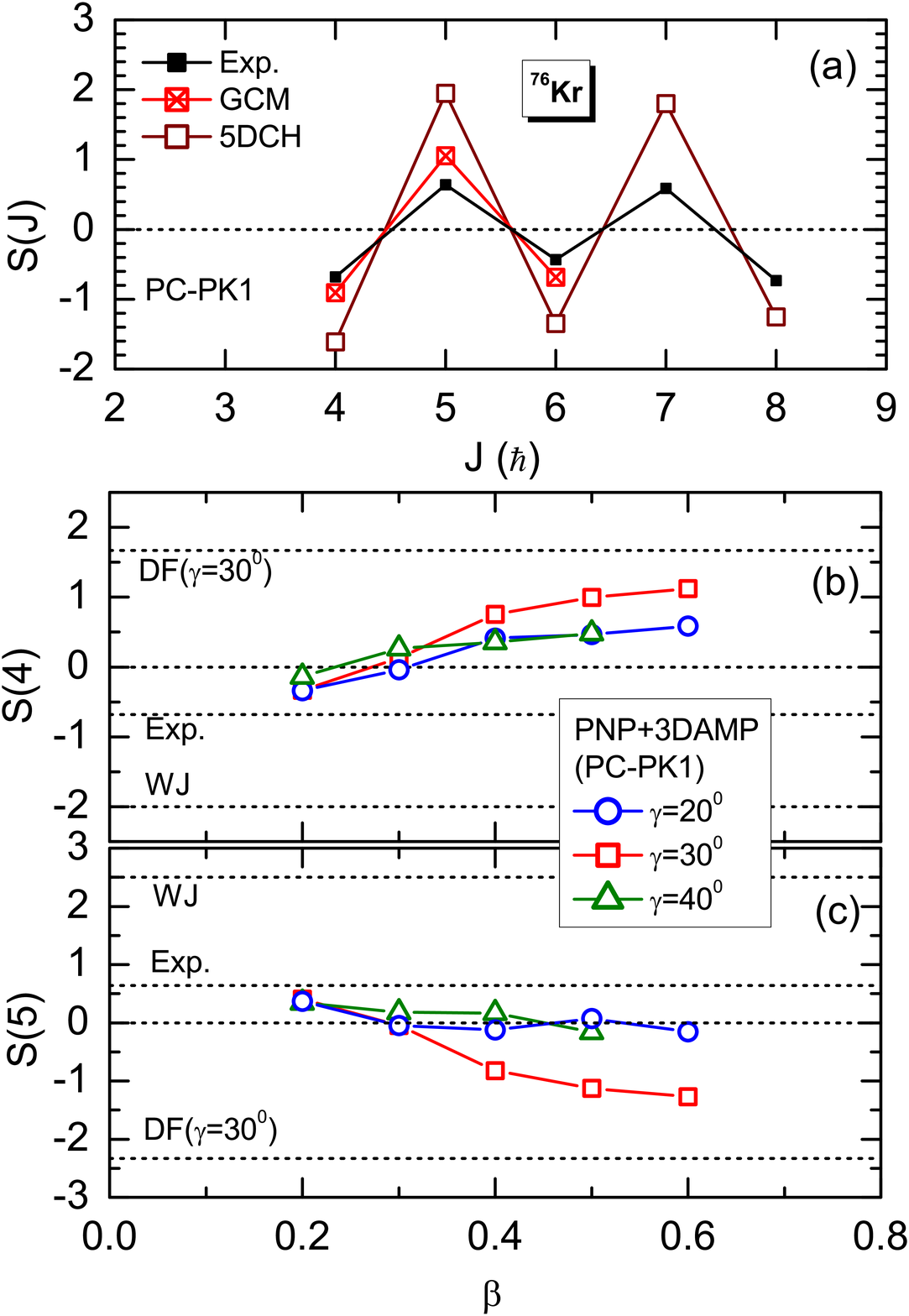}
\caption{\label{staggering}(Color online) Staggering behavior of $\gamma$-band in $^{76}$Kr from the full GCM and 5DCH  calculation, in comparison with the data. The $S(4)$ and $S(5)$ values from the DF ($\gamma=30^\circ$) and WJ models are indicated with horizontal dashed lines.}
\end{figure}

Fig.~\ref{staggering} displays the staggering behavior
\begin{equation}
S(J)=[E(J)+E(J-2)-2E(J-1)]/ E(2^+_1)
\end{equation}
of the odd- and even-spin levels in the quasi-$\gamma$-band. Both the full GCM and the 5DCH calculations reproduce the experimental staggering behavior. However, the dominant configuration of the ground state is different in these two calculations. In the 5DCH results the nearly prolate configurations dominate in the ground state (see Fig.~\ref{gsb} and Ref.~\cite{FU-Y2013_PRC87-054305}), while for the full GCM calculation the triaxial configurations dominate in the ground state. Therefore, it is interesting to know whether the staggering phase provides a reliable fingerprint of rigid triaxiality at low-energies. To address this question, we carry out a PNP+3DAMP calculation based on a fixed configuration with $\beta$-values increasing from 0.2 to 0.6 and with $\gamma=20^\circ, 30^\circ$, and $40^\circ$ respectively.  In Fig.~\ref{staggering} it is shown that the sign of $S(4)$ and $S(5)$ can be inverted if the configuration is changing from $\beta=0.2$ to $\beta=0.4$. The size of $S(4)$ increases with the $\beta$ value after the inversion due to the decreasing of the energy $E(2^+_1)$.  We note that $S(6)$ follows a similar behavior as $S(4)$. This implies that the staggering phase is deformation-dependent. The realistic case is of course much more complicated since all the configurations with different deformations are mixed in the GCM or 5DCH wave functions. Similar phenomena were also observed independently in the most recent triaxial projected shell model (PSM) calculations for $^{76}$Ge~\cite{Bhat14}. In general, the dominated configuration of the low-lying state could change dramatically as a function of angular momentum and as a consequence we will observe a change of the staggering phase of $S(J)$. By this reason, it is not safe to just take the staggering phase of the $\gamma$-band as a signature for rigid triaxiality in low-energy states of realistic nuclei. A clear fingerprint for rigid triaxiality in low-lying states requires further investigation.

\section{\label{sec:conclusion}Conclusions}
In this work we have established a state-of-the-art beyond relativistic mean field method that incorporates the full generator coordinate method together with the techniques of particle number and 3D angular momentum projection. This completely microscopic method has been subsequently applied for studying the triaxiality in the low-lying states of $^{76}$Kr. The low-energy structure has been reproduced very well provided that the triaxiality is taken into account properly. This work provides a complete microscopic study of triaxiality in $^{76}$Kr and, for the first time, a benchmark for the 5DCH based on nuclear EDFs. We have made a detailed comparison between full GCM and 5DCH calculations based on the same EDF. The EDF mapped 5DCH turns out to give very close results to the full GCM calculations, except for an overall overestimation of the excitation energies by about 20\%. The staggering phase of $\gamma$-band has been found to be configuration-dependent and therefore may not be safe to be taken as a signature of rigid triaxiality at low-energy structure. It would be very interesting to repeat such studies with similar calculations based on non-relativistic EDFs~\cite{Bender2008_PRC78-024309,Rodriguez2010_PRC81-064323} and to see, whether they confirm our conclusions.

\begin{acknowledgements}
We thank K. Matsuyanagi, T. Koike, G. F. Bertsch and H. Sagawa for helpful discussions. We also thank the Yukawa Institute for Theoretical Physics at Kyoto University, where this work was completed during the ``International Molecular-type workshop on New correlations in exotic nuclei and advances of theoretical models". This work was supported in part by the Tohoku University Focused Research Project ``Understanding the origins for matters in universe", the Major State 973 Program 2013CB834400, the JSPS KAKENHI Grant Number 2640263, the NSFC under Grant Nos. 11175002, 11105110,  11105111, 11335002, and 11305134, the Research Fund for the Doctoral Program of Higher Education under Grant No. 20110001110087, the Natural Science Foundation of Chongqing cstc2011jjA0376,
the Fundamental Research Funds for the Central Universities (XDJK2010B007 and XDJK2011B002), and from the DFG Cluster of Excellence \textquotedblleft Origin and Structure of the Universe\textquotedblright\ (www.universe-cluster.de)
\end{acknowledgements}

\end{document}